\documentclass[twocolumn,showpacs,preprintnumbers,amsmath,amssymb,prb]{revtex4} 
\usepackage{graphicx}
\usepackage{dcolumn}
\usepackage{bm}
\usepackage{epstopdf}
\usepackage{multirow}

\begin{document}

\title{
Density functional theory study of
phase stability, vibrational and electronic properties of
Mo$_3$Al$_2$C} 
\author{D. Reith}
\email{david.reith@univie.ac.at}
\author{C. Blaas-Schenner}
\author{R. Podloucky}
\affiliation{
Department of Physical Chemistry, University of Vienna and Center for
Computational Materials Science, Sensengasse 8, A-1090 Vienna, Austria
}

\date{\today}

\begin{abstract}
Based on density functional theory the noncentrosymmetric superconductor
Mo$_3$Al$_2$C in its well established $\beta$-Mn type (P4$_1$32) crystal
structure is investigated.  In particular, its thermodynamical and dynamical
stabilities are studied by calculating lattice vibrations within the harmonic
approximation.  It is found that the fully stoichiometric compound is
dynamically unstable.  However, compounds with carbon vacancies, i.e.,
Mo$_3$Al$_2$C$_{1-x}$, can be dynamically stabilized for vacancy concentrations
$x > 0.09$.  By means of a simple thermodynamical model we estimate  $x\sim
0.13-0.14$ for Mo$_3$Al$_2$C$_{1-x}$ at the experimental preparation
temperatures. The influence of the carbon vacancy concentration on the
electronic structure is investigated.
\end{abstract}
\maketitle

\section{Introduction}
Mo$_3$Al$_2$C has been already synthesized in the 1960s \cite{Jeitschko1963}
and subsequently classified as a
superconductor.\cite{Johnston1964,Fink1965,Toth1968,Toth1971} Recently it has
attracted renewed
attention,\cite{Bauer2010,Karki2010,Bonalde2011,Koyama2011,Kuo2012} because its
cubic $\beta$-Mn type crystal structure does not contain a center of inversion.
In such noncentrosymmetric superconductors, as first discussed for
CePt$_3$Si,\cite{Bauer2004} the electron pairing process  is described to be 
be a mixture of spin-singlet and spin-triplet states.
\cite{Gorkov2001,Sigrist} The question whether Mo$_3$Al$_2$C can be classified
as a {\em conventional} or {\em unconventional} superconductor remains still
unresolved according to very recent
investigations.\cite{Bauer2010,Karki2010,Bonalde2011} 
In this work, we will not tackle this issue directly
but will provide results of extensive density functional theory (DFT) calculations on the 
thermodynamical and dynamical stability as well as the electronic structure of this compound 
as a function of carbon content.

Mo$_3$Al$_2$C crystalizes in the cubic $\beta$-Mn type P4$_1$32 structure
containing 24 atoms in the unit cell with, namely 12 Mo, 8  Al, and 4 C atoms.
The C atoms are in the center of regular Mo$_6$ octahedrons which are tilted
to each other.\cite{Jeitschko1963,Bauer2010} The comparison of our DFT
structural parameters to experimental values \cite{Bauer2010} in Table
\ref{tab0} shows both in excellent agreement.

\begin{table}
\caption{\label{tab0} Structural parameters and Wyckoff positions of Mo$_3$Al$_2$C.}
\begin{center}
\begin{tabular}{ l d  d}
\hline
\hline
&    \multicolumn{1}{c}{\quad exp.\footnote{experimental results from Ref. \onlinecite{Bauer2010}}} &\multicolumn{1}{c}{\quad DFT\footnote{DFT calculation (present work)}} \\
\hline
lattice parameter $a$:   &                               6.863\,\text{\AA} & 6.890\,\text{\AA}\\
\hline
Mo on 12d \hfill y:& 0.2025(2)& 0.20183\\
Al on 8c  \hfill x:& 0.068(1)& 0.06658\\
C on 4a           &         &\\
\hline
crystal structure:   & \multicolumn{2}{l}{ \qquad\quad cubic $\beta$-Mn type}              \\
space group:         & \multicolumn{2}{l}{ \qquad\quad 213 or P4$_1$32}                               \\
\hline
\hline
\end{tabular}
\end{center}
\end{table}

\section{Computational Details}

The DFT calculations were done using the Vienna \emph{ab initio} simulation
package (VASP) \cite{Kresse1996,Kresse1999} utilizing  the pseudopotential
construction according to the projector augmented wave
method.\cite{Blochl:1994p8789} For the exchange-correlation functional the
generalized gradient approximation as parametrized by Perdew, Burke, and
Ernzerhof~\cite{PBE1996} was chosen.  The potential for Mo contained 9 valence
states including the 4p semicore states, whereas for Al and C three and four
valence states were included, respectively.  The size of the basis set was
defined by an energy cutoff of 500 eV.  The Brillouin-zone integration  for the
computation of total energies was made using the tetrahedron method with with
Bl\"ochl's corrections \cite{Blochl:1994p13006} based on a $13\times 13\times
13$  Monkhorst and Pack \cite{Monkhorst1976} $\vec{k}$-point mesh, whereas for
the structural relaxations and for the derivation of  the force constants the
first order Methfessel-Paxton smearing method \cite{Methfessel:1989p13056} on a
$7\times 7\times 7$ $\vec{k}$-point mesh was chosen.

The vibrational properties were calculated within the harmonic approximation
by making use of the direct force-constant method as implemented in our
program package fPHON (full-symmetry PHON), which is based on the package
PHON.\cite{Alfe2009}  The structural parameters, i.e., the volume and shape of
the unit cell as well as the positions of the atoms within the unit cell, were
relaxed until the residual forces were less than $10^{-4}$ eV/\AA.
Furthermore, for a high accuracy of the phonon calculations the force
constants derived from the displaced atoms were corrected by subtracting the
tiny but still finite forces of the fully relaxed structures. Some of the
phonon dispersions were cross checked by  using density functional
perturbation theory~\cite{baroni2001} (DFPT) as implemented in VASP.

\section{Vibrational Properties} 
\label{phonons}

\begin{figure*}
 \begin{center}
\begin{tabular}{  c   c}
 (a) &  (b)\\
\includegraphics[width = 0.385\textwidth ]{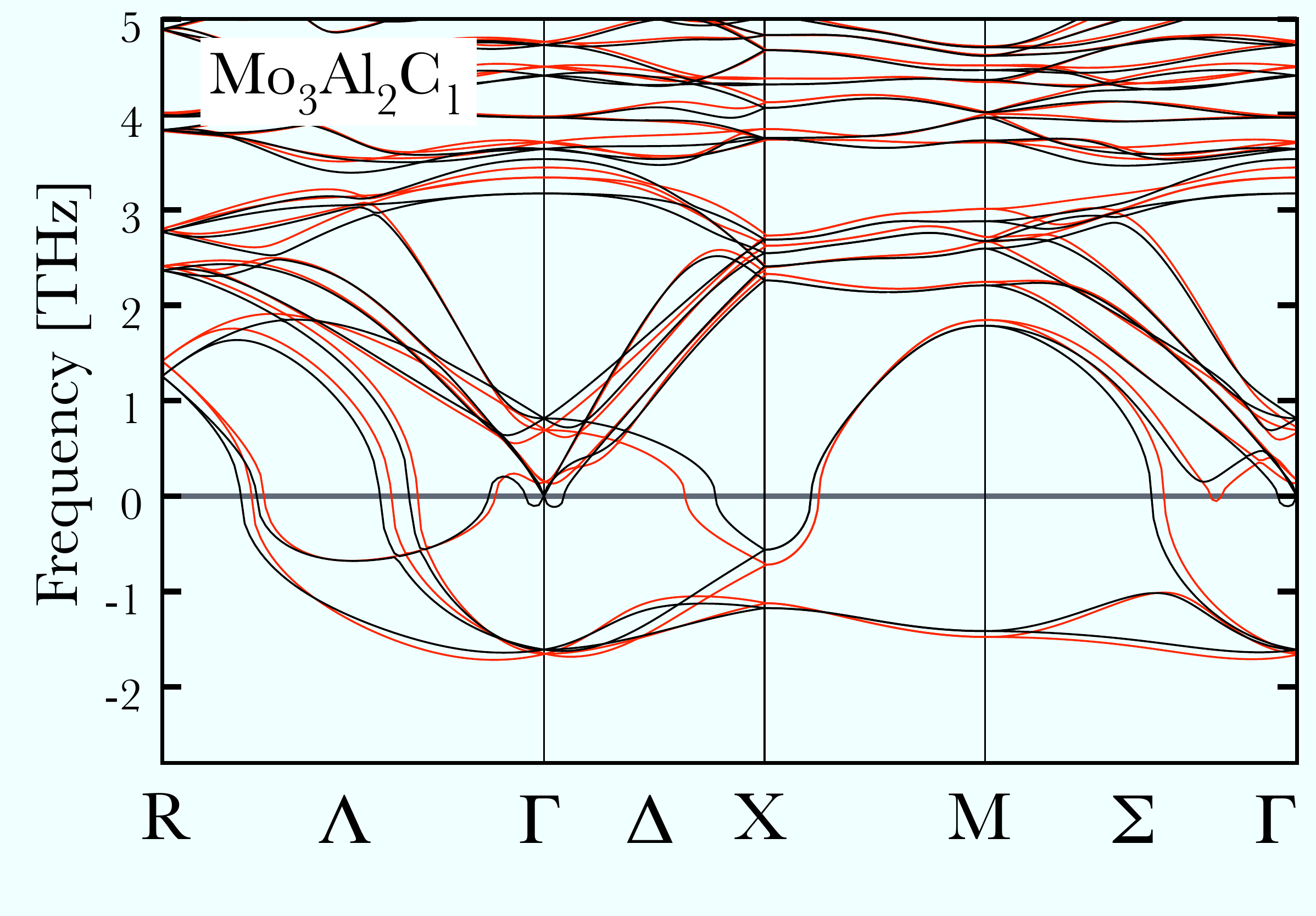}&
\includegraphics[width = 0.585\textwidth ]{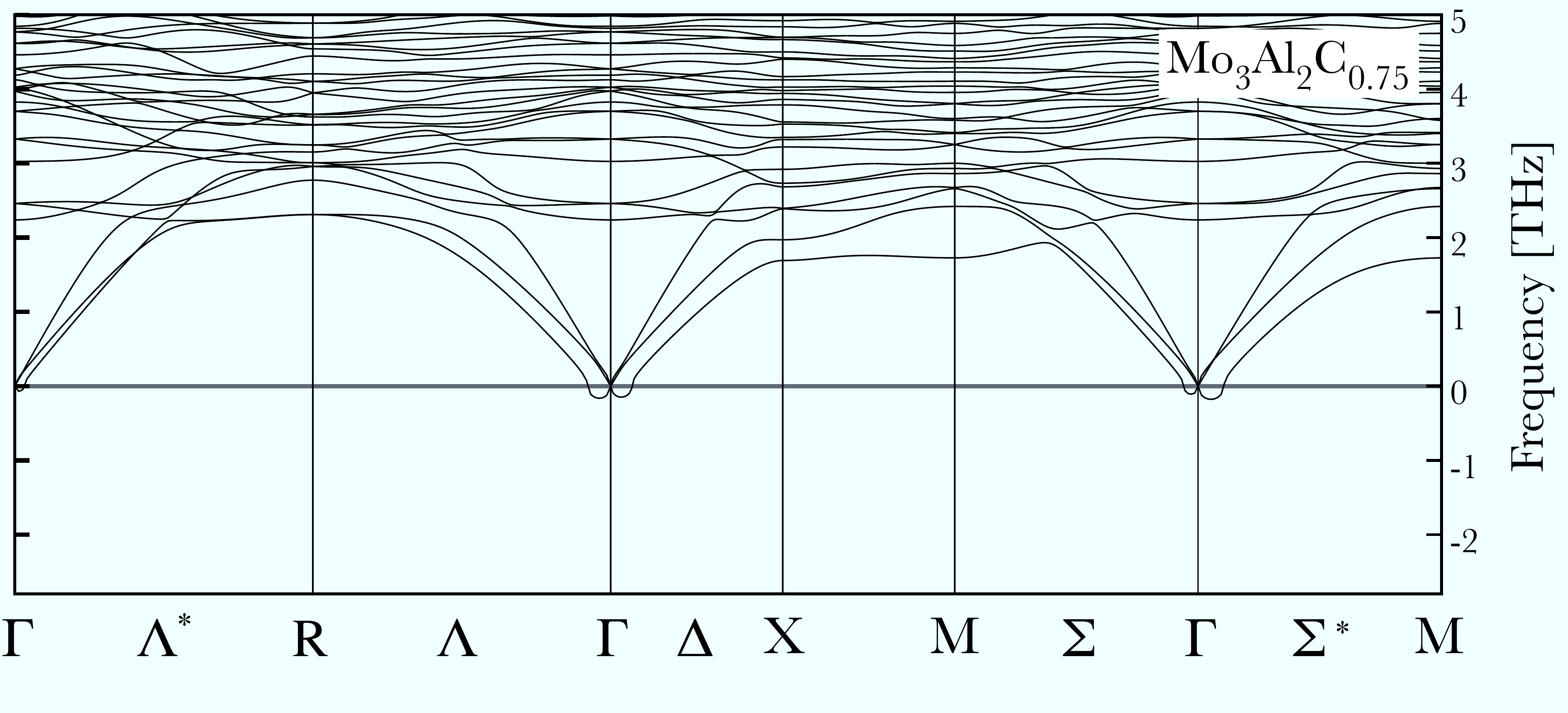}\\
\end{tabular} 
\end{center}
\caption{(Color online) Phonon dispersions up to 5 THz of 
stoichiometric Mo$_3$Al$_2$C (a) and of Mo$_3$Al$_2$C$_{0.75}$ (b).  
Imaginary frequencies are shown as negative values.
For Mo$_3$Al$_2$C in panel (a) the calculated phonon dispersion
derived from the direct force-constant method (black lines) is compared to
results from DFPT theory (red lines). 
For Mo$_{3}$Al$_{2}$C$_{0.75}$ in panel (b) the dispersion
relation as calculated from  the direct force-constants method (black lines) is presented. 
It should be noted that because of the reduced symmetry due to the carbon
vacancy the dispersions along the paths  
 $\Lambda^*=\pm[\xi,-\xi,-\xi]$ and  $\Sigma^*=\{\pm[\xi,\xi,0], \pm
[\xi,0,\xi], \pm[0,-\xi,\xi] \}$ with $0<\xi<\pi/a$ differ to the other symmetry related
$\vec{k}$-directions refered by
$\Lambda$ and $\Sigma$, respectively.
}
\label{fig1}
\end{figure*}

It turns out that perfectly stoichiometric Mo$_3$Al$_2$C is dynamically
unstable.  According to panel (a) of Fig.  \ref{fig1}  optical modes with
imaginary frequencies around $\Gamma$ arise. The dynamical instability seems
surprising considering the well-established crystal structure of
Mo$_3$Al$_2$C.\cite{Jeitschko1963,Johnston1964,Fink1965,Toth1968,Toth1971,Bauer2010,Karki2010,Bonalde2011,Koyama2011,Kuo2012}
One should, however, be aware that the actual carbon content is
experimentally difficult to discern by X-ray diffraction because of carbons
comparatively small X-ray cross section\cite{McMaster1969} and therefore
there exists some uncertainty with respect to carbon vacancies. In general, many
carbides are prone to have vacancies on the C sublattice. 

Before speculating on the physical explanation of the detected instability
we undertook numerical and methodological tests. First of all, it is assured
that the results are converged with respect to the number of $\vec{k}$-points.
Furthermore, 
for deriving the force
constants several atomic displacements of the atoms, i.e., $|\vec{u}|=\{0.01,
0.02,0.05, 0.1, 0.2,0.3,0.4\}~\text{\AA}$ were chosen and  
calculations without any symmetry  were performed.  All 
calculations confirm the existence of imaginary modes
for Mo$_3$Al$_2$C.
In particular, the calculation with the smallest displacement of $|\vec{u}|=0.01~\text{\AA}$
yielded imaginary optical modes with a frequency of $1.62~i~$THz  at $\Gamma$. 
As a further test the DFPT~\cite{baroni2001} technique as
implemented in VASP was applied also resulting in imaginary modes of $1.65~i~$THz at $\Gamma$. 
From these DFPT calculations force constants have been derived
and used as input for \emph{f}PHON. Both, the direct force-constant method
and the DFPT treatment gave very similar results for the phonon
dispersions as shown  in panels (a) of Fig.  \ref{fig1}.  

In order to study if another similar structure exists that is dynamically
stable and energetically more favorable compared to the cubic $\beta$-Mn
type structure, all atoms were displaced from their equilibrium
positions. In addition, a tetragonal deformation was enforced onto the unit cell.
Using VASP this deformed structure was subsequently relaxed without any symmetry
constrains, resulting in the well-known crystal structure described
before. 

Because the perfectly stoichiometric compound Mo$_3$Al$_2$C is found to be
dynamically unstable, we investigated if dynamical stabilization can be
achieved by vacancies, in particular by carbon vacancies.

In general, vacancies on all sublattices will stabilize the phonon dispersion,
at least above a certain concentration.
Vacancies on the Mo or Al
sublattice are less likely to exist since they would be easily
detectable, i.e., they have a comparatively large X-ray cross section.\cite{McMaster1969}  
Furthermore, our DFT derived vacancy formation energies in the following
Section~\ref{formation} strongly indicate that Mo or Al vacancies are
thermodynamically too costly to be formed.

Assuming a certain carbon vacancy concentration suitable supercell
calculations were done  for the defect structures. 
Panel (b) in Fig. \ref{fig1} shows the phonon dispersion for Mo$_3$Al$_2$C$_{0.75}$, i.e., for a single
carbon vacancy in the standard unit cell ($1\times 1\times 1$ supercell) with
3 out of possible 4 carbon sites occupied. 
Further calculations,
of which the  dispersions are not shown, were also done for a
single carbon vacancy in a $2\times 1\times 1$ supercell 
(Mo$_{3}$Al$_{2}$C$_{0.875}$ with 7 out of possible 8 C
sites occupied) and in a $2\times 2\times 2$ supercell
(Mo$_{3}$Al$_{2}$C$_{0.96875}$ with 31 out of 32 C sites occupied). 
Both Mo$_{3}$Al$_{2}$C$_{0.75}$
and Mo$_{3}$Al$_{2}$C$_{0.875}$ are found to be dynamically stable with no
imaginary modes, whereas 
Mo$_{3}$Al$_{2}$C$_{0.96875}$ is found to be significantly unstable with the
value of the lowest imaginary optical modes at $\Gamma$ being $1.26~i~$THz
quite close to that of the perfect stoichiometric compound ($1.62~i~$THz). 

\begin{figure*}
 \begin{center}
\begin{tabular}{c c}
(a) & (b)\\
\includegraphics[width = 0.49\textwidth ]{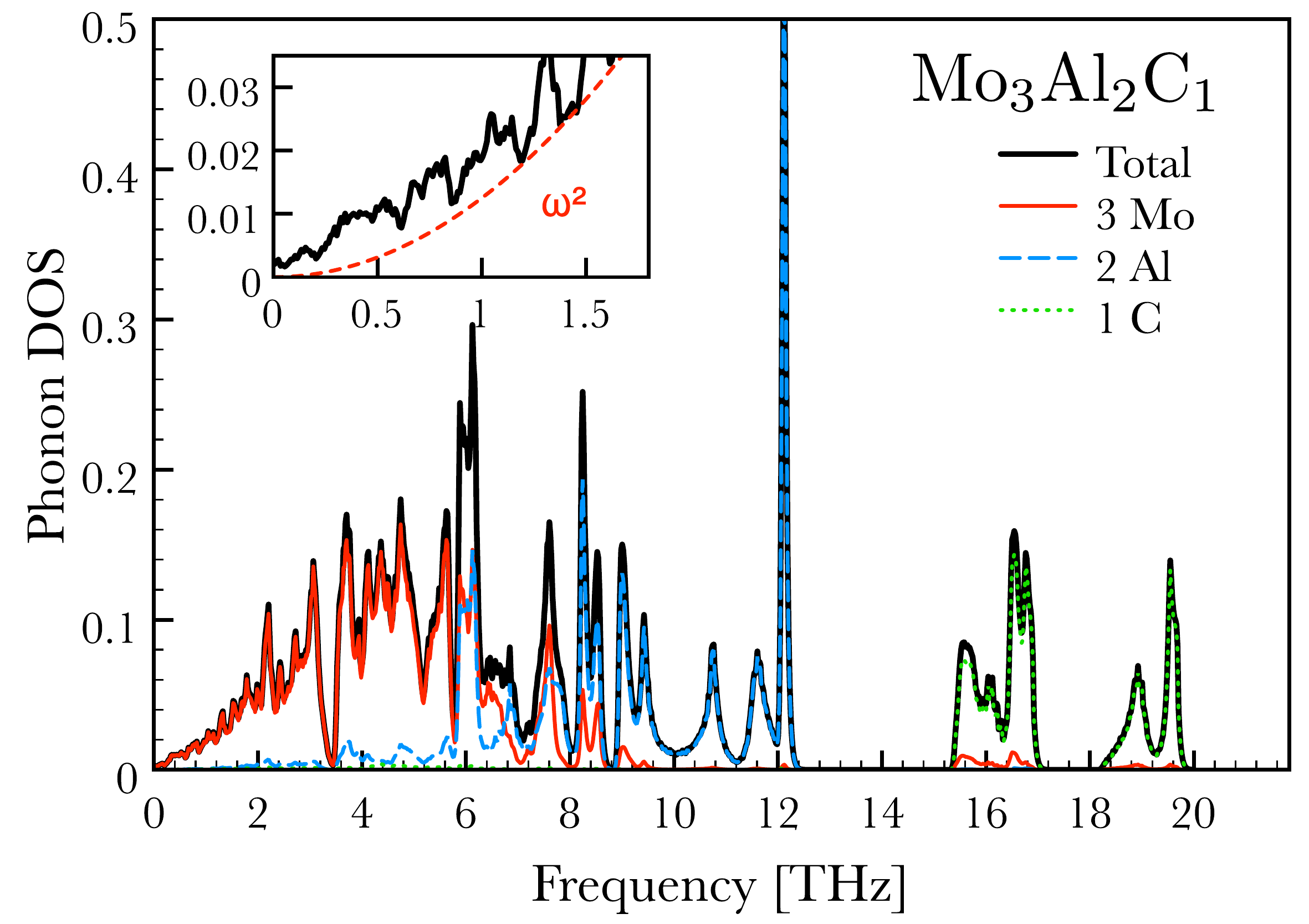} &
\includegraphics[width = 0.49\textwidth ]{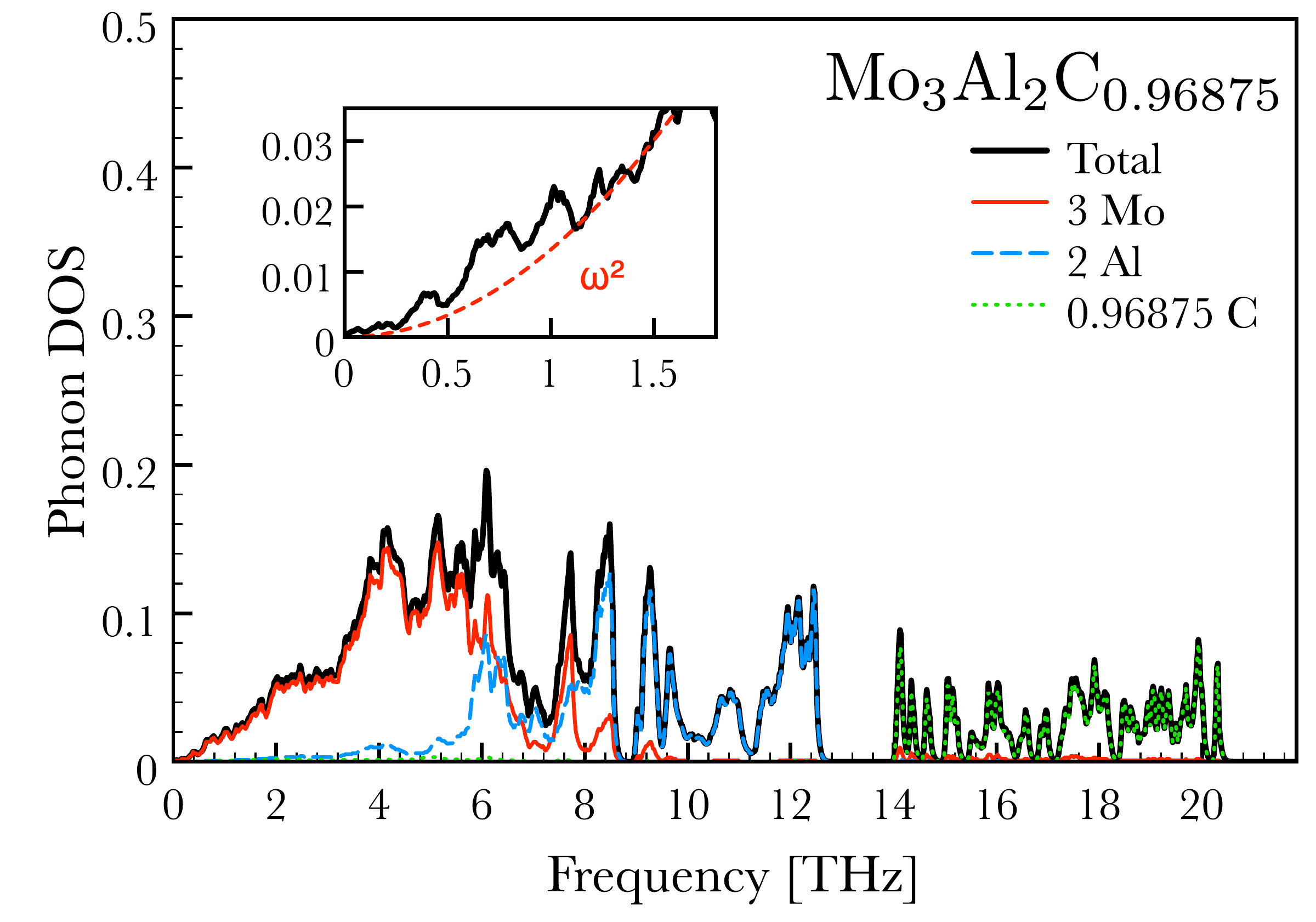}\\
&\\
&\\
(c) & (d)\\
\includegraphics[width = 0.49\textwidth ]{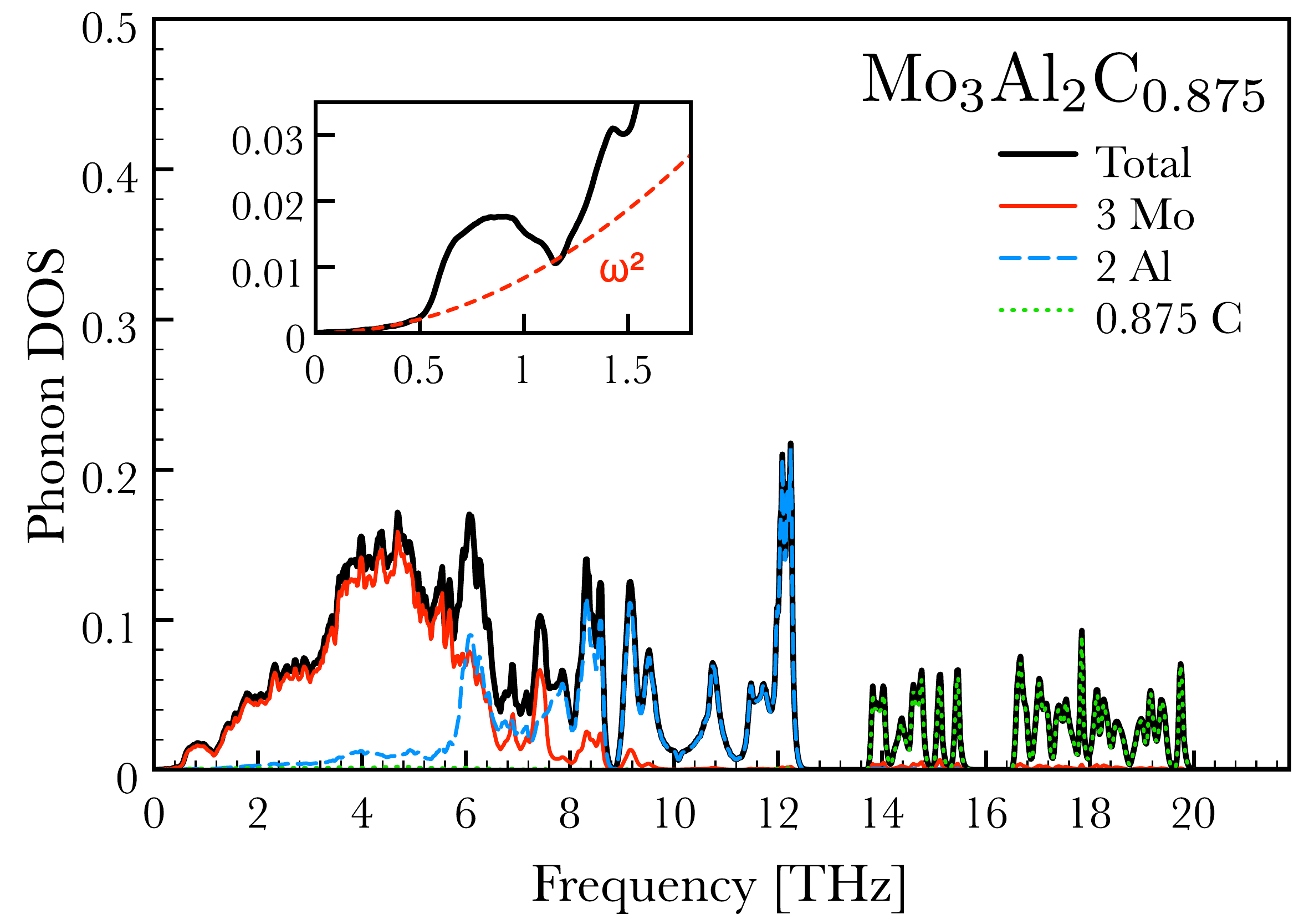}&
\includegraphics[width = 0.49\textwidth ]{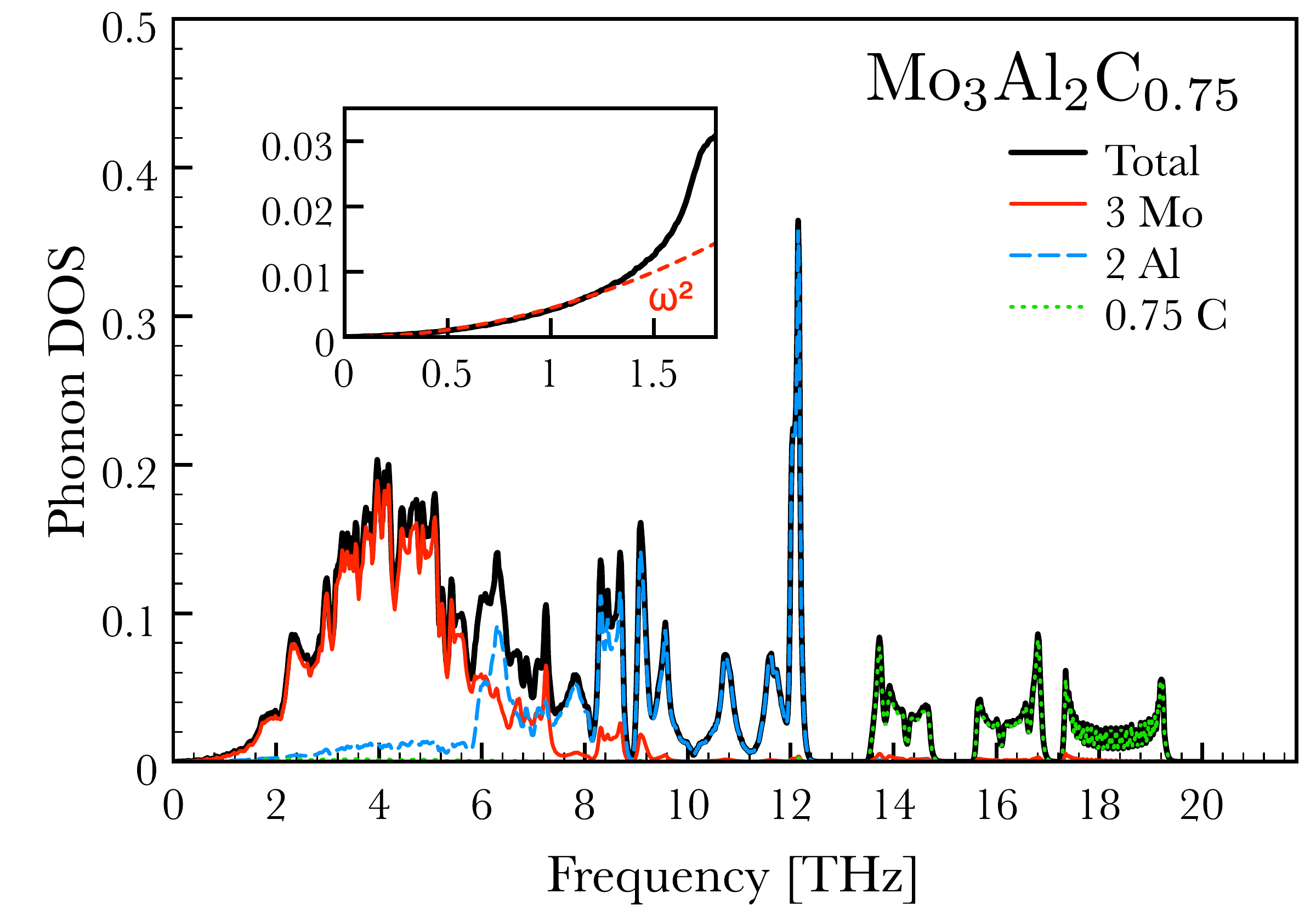} \\
\end{tabular}
\end{center}
\caption{
(Color online) Total and partial phonon DOS of  Mo$_{3}$Al$_{2}$C$_1$ (a), Mo$_{3}$Al$_{2}$C$_{0.96875}$ (b),
Mo$_{3}$Al$_{2}$C$_{0.875}$ (c), and Mo$_{3}$Al$_{2}$C$_{0.75}$ (d) on the
Total phonon DOS (black solid line), partial  DOS of Mo (red solid line),
partial DOS of Al (blue dashed line), and partial DOS of C (green dotted
line). In the insets the total DOS at low frequencies is compared to a
Debye-like $\omega^2$ behavior (red dashed line).     
}
\label{fig3}
\end{figure*}

In Fig.  \ref{fig3} the normalized phonon density of states (DOS) of the
dynamically unstable compounds
Mo$_{3}$Al$_{2}$C$_{1}$ and Mo$_{3}$Al$_{2}$C$_{0.96875}$ are compared to 
those of 
the  dynamically
stable compounds Mo$_{3}$Al$_{2}$C$_{0.875}$ and Mo$_{3}$Al$_{2}$C$_{0.75}$.
While at low frequencies no Debye-like $\omega^2$ behavior is observed 
for Mo$_{3}$Al$_{2}$C$_1$ and  Mo$_{3}$Al$_{2}$C$_{0.96875}$, 
it is seen for the other two cases. For
Mo$_{3}$Al$_{2}$C$_{0.875}$ the Debye-like behavior is observed only in a
rather narrow frequency range up to $0.5~$THz
due to the softening of the acoustical and optical modes. However, for Mo$_{3}$Al$_{2}$C$_{0.75}$ the
Debye-feature reaches up to $1.4~$THz.  The partial DOS
reveals that  Mo, as the heaviest atomic species, dominates the lower
frequency spectrum
(up to $7.5~$THz), whereas C, being the lightest atom,
has contributions only at frequencies above $13.5~$THz. 
Furthermore, the carbon dominated frequency modes
are shifted down by the introduction of vacancies from above $15~$THz for Mo$_{3}$Al$_{2}$C$_{1}$  to about
$13.5~$THz for Mo$_{3}$Al$_{2}$C$_{0.75}$.
Strikingly, in all the
DOS's a pronounced Al peak at $\approx12.2~$THz occurs.
 The Al spectrum is rather broad with significant contributions from $6~$THz to $12.2~$THz.
Even below this range a telling small contribution is found indicating a
hybridization with Mo modes.

\begin{figure}
 \begin{center}
\includegraphics[width = 0.4\textwidth ]{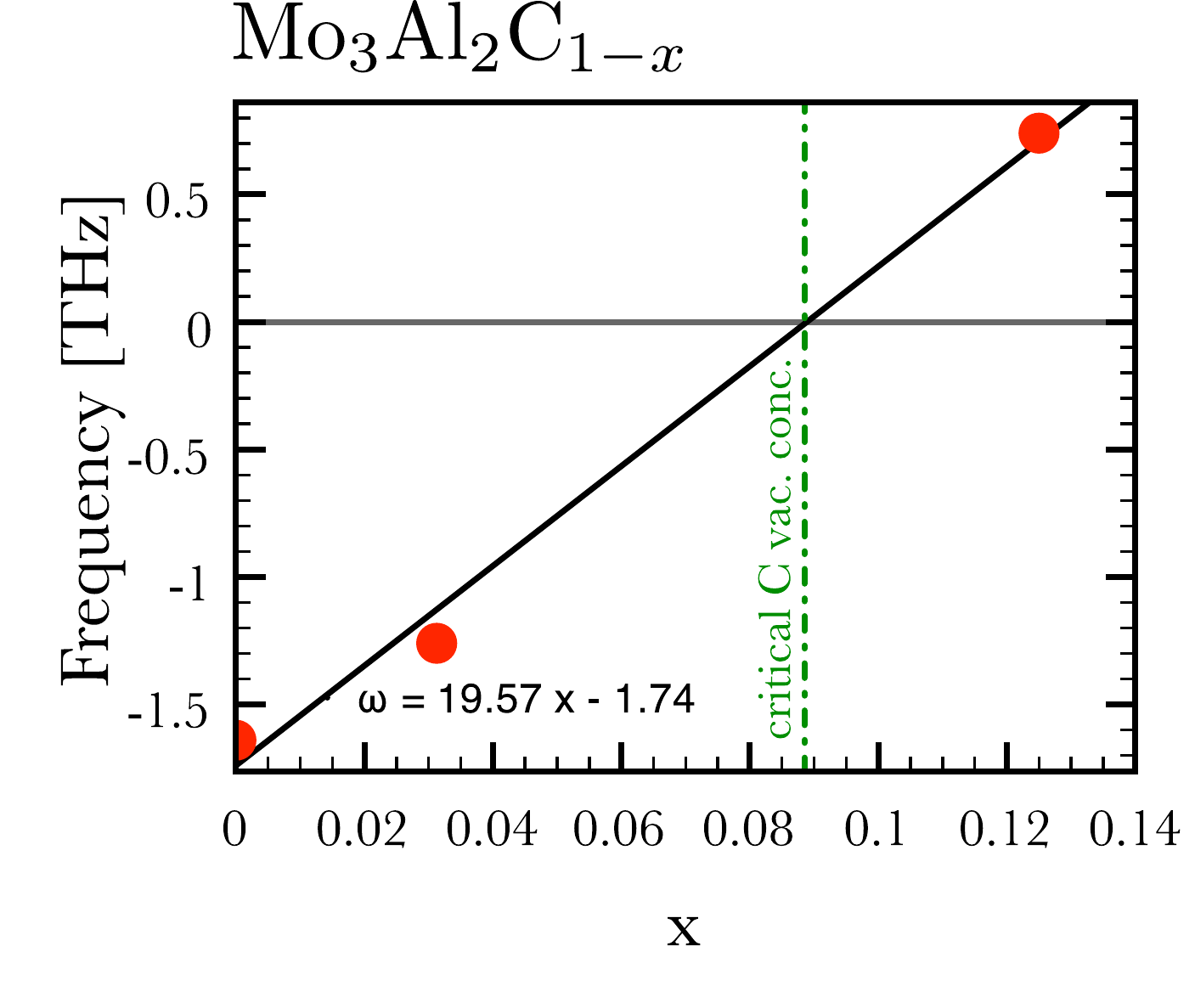} 
 \end{center}
\caption{(Color online) Frequency of the lowest optical mode at $\Gamma$ versus carbon vacancy concentration $x$ 
as calculated (red circles) and linearly interpolated (solid black line). 
The predicted critical carbon vacancy concentration is indicated (green dashed-dotted line).}
\label{fig2}
\end{figure}

The results indicate, that there is a critical concentration of carbon
vacancies below which the compound becomes dynamically unstable. Assuming
that the frequency of the lowest optical mode at $\Gamma$ scales linearly with
the carbon vacancy concentration $x$, the critical carbon vacancy concentration is estimated
to be $x_\text{crit}\sim0.09$ 
(using the values of Mo$_{3}$Al$_{2}$C$_{1-x}$ at $x=0,$ $0.03125$, and $0.125$
as input; see Fig. \ref{fig2}).

The cause of the stabilization of the optical low-frequency Mo modes by
the carbon vacancies is the changed Mo-C bonding in the Mo$_6$C
subunits. We examined the relaxations occurring in  Mo$_{3}$Al$_{2}$C$_{0.75}$,
i.e., when a single C atom is removed from one of the four Mo$_6$C subunits in
the unit cell of Mo$_{3}$Al$_{2}$C$_{1}$.  Thereby a strong influence on the
Mo-C bonds in all three remaining Mo$_6$C subunits is observed because they
share a common Mo atom with the defect subunit.  By relaxing the atomic
positions the C atom drifts into an off-center position within the remaining
Mo$_6$C subunits, increasing the average Mo-C bond length by 1.4 \% and the
corresponding octahedral volume by 3.6 \%.  The distortion of the remaining
Mo$_6$C subunits seems to be the stabilizing factor for the vibrational modes.
Concomitantly we notice this distortion for the Mo-Al bonding, e.g., the two
distinctive nearest Mo-Al bonds with bond-lengths 2.84~\AA~and 2.95~\AA~in the
fully stoichiometric compound are distorted in Mo$_{3}$Al$_{2}$C$_{0.75}$ to
lengths in the range of $2.79-3.01~$\AA.  Such a distortion for the Mo-Al bonds
has recently been indicated experimentally.\cite{Kuo2012}

\section{Carbon Vacancies}
\label{formation}

As discussed, a certain amount of carbon vacancies is needed to
stabilize the imaginary optical modes. The key question is
if the formation of vacancies is at all thermodynamically possible. 
This question is investigated by calculating vacancy formation energies and
by means of a model.

Within a standard DFT approach, the vacancy formation energy
$\varepsilon^{X}_\text{vac}$ per atom $X$ is defined by
subtracting the total energy $E_\text{DFT}(\text{Mo}_{12}\text{Al}_8\text{C}_4)$ of the
stoichiometric compound from the total energy $E_\text{DFT}(\text{Mo}_{12}\text{Al}_8\text{C}_4-X)$ of
the compound with a vacancy of atom type $X$ and adding the ground-state energy 
$E_\text{DFT}(X)$ of the removed atom $X$ by
\begin{eqnarray} 
\label{eq:epsilon}
\varepsilon^{X}_\text{vac}&=&E_\text{DFT}(\text{Mo}_{12}\text{Al}_8\text{C}_4-X)+E_\text{DFT}(X)-
\nonumber\\&& 
-E_\text{DFT}(\text{Mo}_{12}\text{Al}_8\text{C}_4)~. 
\end{eqnarray}
Because standard DFT calculations are strictly valid only at $T=0~$K
no temperature dependency has been yet introduced. This can be done by
considering  the temperature dependent
vibrational free energies $F_\text{vib}$ and defining the vibrational vacancy
formation energy per $X$ atom  similar to
Equ.~\ref{eq:epsilon},\cite{reith2009}
\begin{eqnarray}
\label{eq:f}
f^{X}_\text{vac}(T)&=&F_\text{vib}(\text{Mo}_{12}\text{Al}_8\text{C}_4-X)+F_\text{vib}(X)-
\nonumber\\&&
-F_\text{vib}(\text{Mo}_{12}\text{Al}_8\text{C}_4)~. 
\end{eqnarray}
Both, $\varepsilon^{X}_\text{vac}$ and $f^{X}_\text{vac}(T)$ are formulated for a standard unit cell. To derive results for
smaller vacancy concentrations larger supercells are needed and the
stoichiometries in Equs.~\ref{eq:epsilon} and \ref{eq:f} have to be scaled accordingly. 
The reference energies $E_\text{DFT}(X)$ and $F_\text{vib}(X)$ were derived 
for the ground states of body-centered cubic Mo, face-centered cubic Al, and C in the graphene structure.

\begin{table}
\caption{\label{tab2} Vacancy formation energies in eV for Mo$_3$Al$_2$C
for $1\times 1\times 1$ and  $2\times 1\times 1$ supercells
without vibrational contributions (DFT, $T=0~$K) and with $f^{X\text{(vac)}}(T)$ at 
 $T=1523~$K and $1773~$K.  }
\begin{center}
\begin{tabular}{l l c c c c }
\hline
\hline
&			              &  Mo-vac              & Al-vac            & C-vac            & C-vac \\

\hline
        \multicolumn{2}{c}{supercell size}              &1$\times$1$\times$1&1$\times$1$\times$1&1$\times$1$\times$1&2$\times$1$\times
$1\\
\hline
&$\varepsilon
$                    &       1.74    &         0.86        &  0.60  & $~~0.56$   \\
$T=1523~$K:&
$\varepsilon+f(T)$&       1.58    &         0.50        &  0.20  & $-0.39$   \\ 
$T=1773~$K:&
$\varepsilon+f(T)$&       1.51    &         0.39        &  0.08   & $-0.66$   \\ 
\hline
\multicolumn{4}{l}{C vacancy concentration $x$ for Mo$_3$Al$_2$C$_{1-x}$:}& 0.25 &$~~0.125$\\
\hline
\hline
\end{tabular}
\end{center}
\end{table}
\begin{figure}
\begin{center}
\includegraphics[width = 0.49\textwidth ]{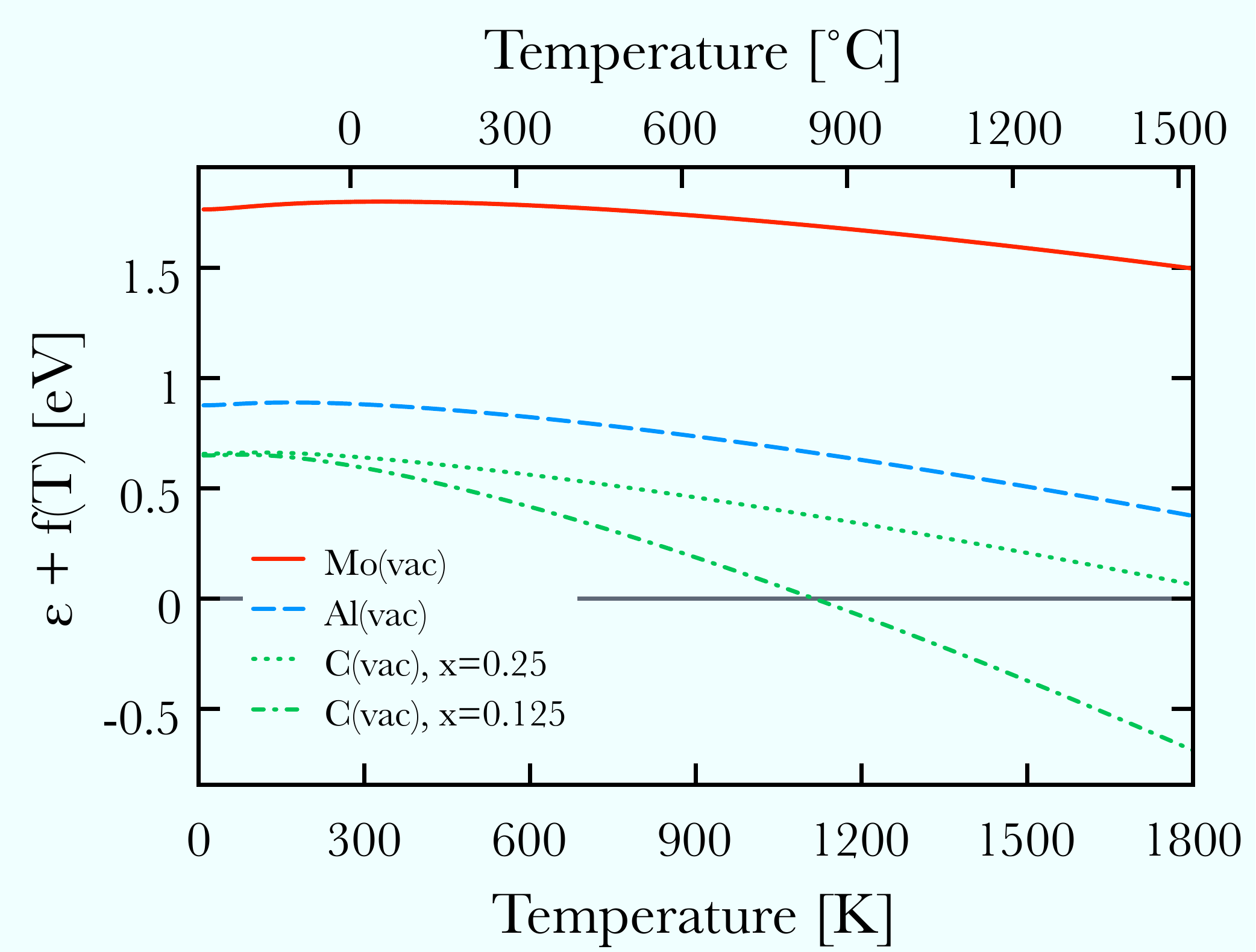}
\end{center}
\caption{(Color online) Vacancy formation energy $\varepsilon +f(T)$ as the sum of the DFT
$\varepsilon$ and vibrational $f(T)$ formation energy  versus temperature for a
Mo vacancy (red solid line), a Al vacancy (blue dashed
line), a C vacancy (green dotted line) in a unit cell
($1\times1\times1$ supercell), and for a C vacancy in a
$2\times1\times1$ supercell (green dashed-dotted line). }
\label{fig4}
\end{figure} 

The vacancy formation energies in Table \ref{tab2} and Fig. \ref{fig4} 
at $T=0~$K 
are strongly positive for all types of vacancies, whereby the Mo vacancy with
its formation
energy of 1.74 eV is by far the most unfavorable one. Carbon
vacancies are the most favorable ones with a formation energy of 0.60 eV
for Mo$_{3}$Al$_{2}$C$_{0.75}$. This value is reduced by only 0.04 eV 
for the smaller vacancy concentration of Mo$_{3}$Al$_{2}$C$_{0.875}$. 

The experimental samples were synthesized
at $T=1773~$K  and heat treated at $1523~$K.\cite{Bauer2010} Therefore, theory
needs to consider 
temperature dependent vacancy formation energies
$f^{X}_\text{vac}(T)$ combined with the composition dependent configurational
entropy $S_\text{conf}(x)$~\cite{reith2009} in order to compare with experiment. 
For the actual calculation of the vibrational free energy the small amount of
imaginary modes in the fully stoichiometric
compound $\text{Mo}_{3}\text{Al}_{2}\text{C}_{1}$ were omitted.

Table \ref{tab2} and Fig. \ref{fig4}  show that
the vibrational contributions reduce the
strongly positive vacancy formation energies at $T=0~$K.
While this reduction is rather small for the Mo vacancy 
(from $1.74~$eV to $1.51~$eV at $1773~$K),
it is much larger for the other two types of vacancies.

In particular, the formation energy of the carbon vacancy in 
Mo$_{3}$Al$_{2}$C$_{0.75}$ decreases from $0.60~$eV to $0.08~$eV
at $1773~$K. Remarkably, this reduction is much larger for the smaller carbon vacancy
concentration, i.e., Mo$_{3}$Al$_{2}$C$_{0.875}$, with a decrease 
by more than $1$ eV down to $-0.66~$eV. 
These negative values for the formation energy indicate a possible thermodynamical 
stabilization of carbon vacancies in  Mo$_{3}$Al$_{2}$C$_{1-x}$. 
Noticeably, this difference of the temperature dependent free energies for
different carbon vacancy concentrations comes exclusively from the vibrational
contributions $f(x,T)$, because --as mentioned above-- 
at $T=0~$K the vacancy formation energies are almost equal.

\begin{figure}
 \begin{center}
\includegraphics[width = 0.49\textwidth ]{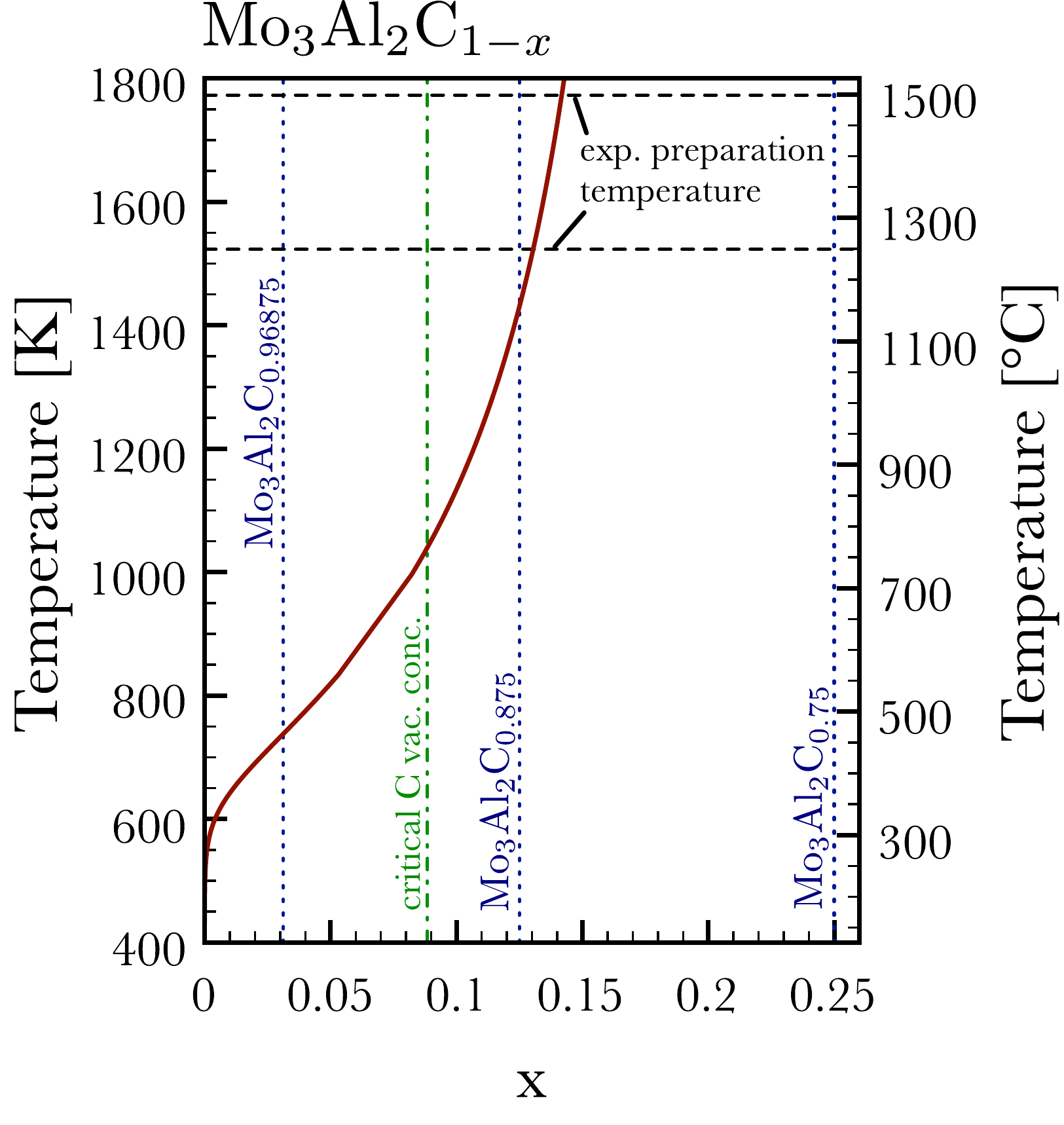} 
 \end{center}
\caption{(Color online) Temperature dependent C vacancy concentration $x(T)$ plotted as a solid
red line. Experimental preparation temperatures are indicated as dashed black lines.
The critical C vacancy concentration, below which Mo$_3$Al$_2$C$_{1-x}$ gets
dynamically unstable, is shown as a green dash-dotted line, while the
C vacancy concentrations of the calculated supercells are drawn as blue dotted lines.     
}
\label{fig5}
\end{figure}

From an isolated defect model~\cite{Ashcroftde} the temperature dependent
equilibrium vacancy concentration $x$ can be calculated.  However, the vacancy
formation energy $\varepsilon(x) + f(x,T)$ in the description of the internal
energy, $U(x,T)=(\varepsilon(x) + f(x,T))x$,\cite{Mayer1997} is strongly
dependent on $x$. Hence, the isolated defect model cannot be applied directly,
as the internal energy  $U(x,T)$ is not a linear function of $x$. In our case
it is described as a quadratic function of $x$, $U(x,T)=ax^2+bx$ wherein $a$
and $b$ are temperature-dependent parameters fitted to the calculated carbon
values at $x=0$, $0.125$, and $0.25$.

Thus, the free energy for the vacancy formation
including the configurational entropy is formulated as
\begin{equation}
F(x,T)=ax^2+bx-k_{B}TS_\text{conf}(x)~.
\end{equation}
Assuming $S_\text{conf}(x)$ is the mixing entropy of non-interacting vacancies
$S_\text{conf}(x)=x\ln(x)+(1-x)\ln(1-x)$, the derivative of $F$ with respect to
$x$ can be used to search for the temperature dependent concentration $x(T)$ by
minimizing the free energy, i.e.,  
\begin{equation}\label{eq:xcrit}
\frac{\partial F(x,T)}{\partial x}=0 \quad \Rightarrow \quad x = \frac{e^{-\beta(2ax+b)}}{1+e^{-\beta(2ax+b)}}~,
\end{equation}
with $\beta=1/(k_B T)$. This expression enables the numerical calculation of $x(T)$, as
shown in Fig. \ref{fig5}, using a bracketing root finding
algorithm.\cite{HPress:2007p15452} Inspecting Fig. \ref{fig5} the C vacancy concentration is 0.13$-$0.14.
at the experimental preparation temperatures~\cite{Bauer2010} of $1773~$K and
$1523~$K. As elaborated in the previous Section~\ref{phonons} at such vacancy
concentrations Mo$_3$Al$_2$C$_{1-x}$ is dynamically stable.

\section{Electronic Structure}

\begin{figure*}
 \begin{center}
\begin{tabular}{  c   c}
 (a) &  (b)\\
\includegraphics[width = 0.40\textwidth ]{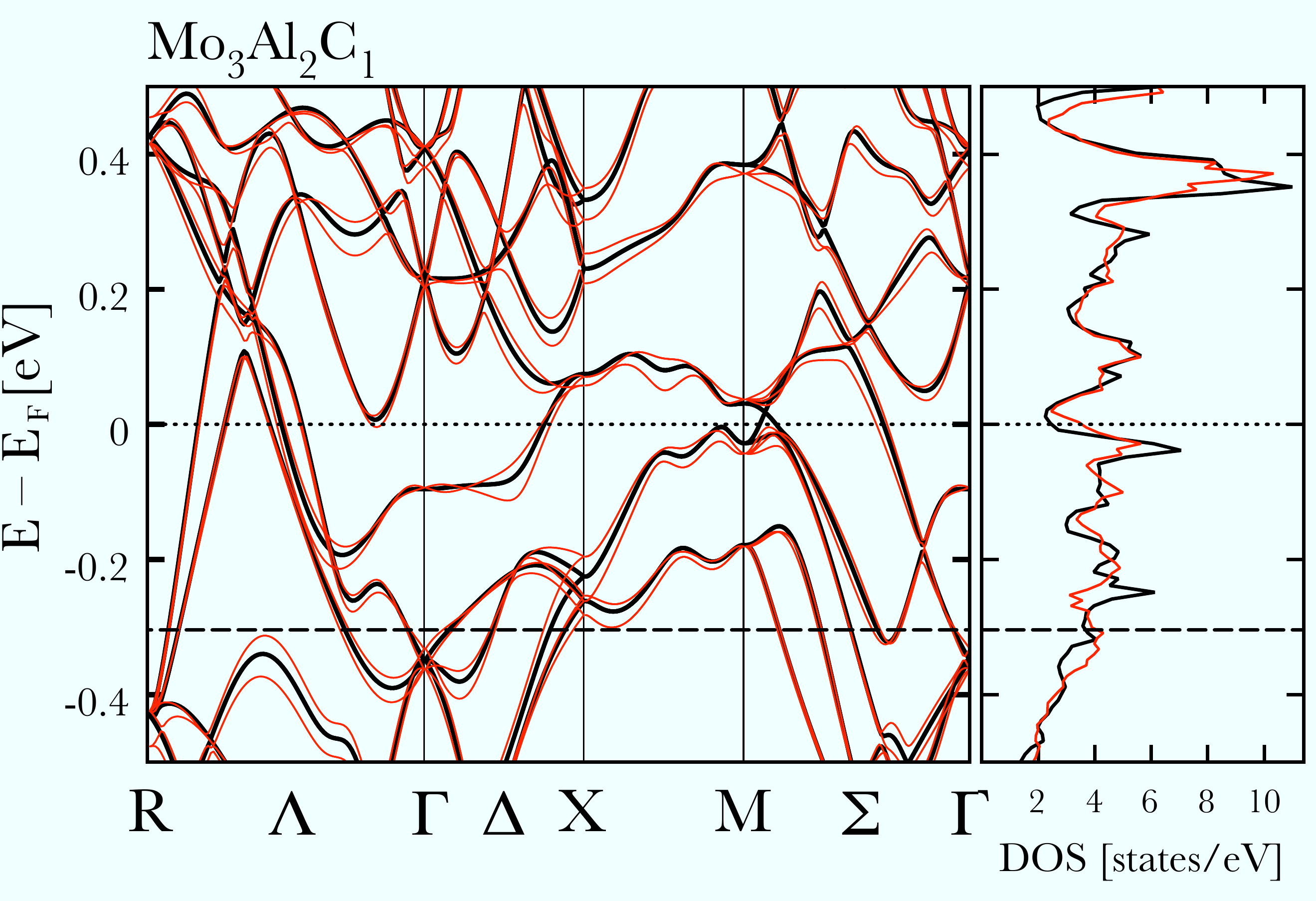}&
\includegraphics[width = 0.555\textwidth ]{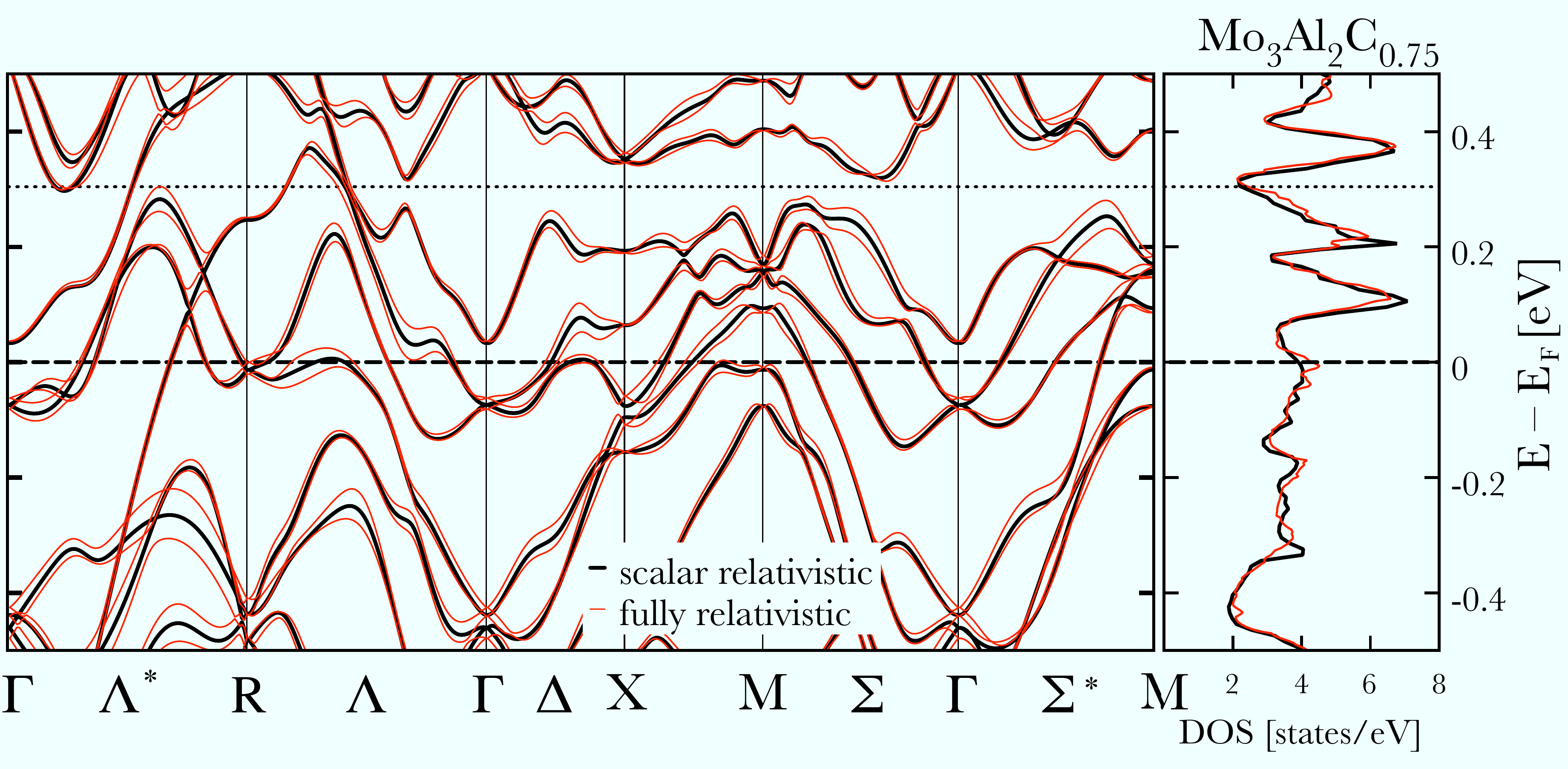}\\
\end{tabular} 
\end{center}
\caption{(Color online)
Electronic band structure and DOS of stoichiometric  Mo$_3$Al$_2$C (a) and of
Mo$_3$Al$_2$C$_{0.75}$ (b) calculated scalar relativistically (black lines) and
fully relativistically including spin-orbit coupling (red lines).  The Fermi
energy $E_\text{F}$ referring to the number of valence electrons of Mo$_{3}$Al$_{2}$C
is indicated as a dotted line while the Fermi energy corresponding to
Mo$_{3}$Al$_{2}$C$_{0.75}$ is indicated as a dashed line.  
Note the different directions $\Sigma, \Sigma^*$ and $\Lambda, \Lambda^*$ as
discussed in the caption of Fig.~\ref{fig1}.
}
\label{fig6}
\end{figure*}

After finding that vacancies do exist on the carbon sublattice and that these 
are necessary to stabilize the crystal structure
we will now briefly discuss the band structure and the electronic DOS of 
Mo$_{3}$Al$_{2}$C$_{1-x}$. As will be shown the influence of a changed carbon
stoichiometry on the band structure can not be described by a simple rigid band model. 
Especially, the spin-orbit splitting on the bands in a fully relativistic
calculation strongly depends on $x$.

In Fig. \ref{fig6} the electronic band structure and DOS of the fully
stoichiometric compound is compared to that of Mo$_{3}$Al$_{2}$C$_{0.75}$.
The attentive reader might question this choice of $x=0.25$ as being too high,
because our thermodynamical model (described in the previous section) predicted
a much lower carbon vacancy concentration. However, we chose this value due to the
fact that the unit cells of both compounds have equal size and shape making it
easier to compare these.  

At this point it should be remarked that our calculated band structure for
Mo$_{3}$Al$_{2}$C$_{1}$ shown in panel (a) in Fig. \ref{fig6} resembles the one
published previously by Bauer \emph{et al.}\cite{Bauer2010} but differs
distinctively from that published by Karki \emph{et al}.\cite{Karki2010} The
different finding by Karki \emph{et al.} can only stem from the use of a
different crystal structure.  We have recalculated the band structure by means
of the full-potential linearized augmented plane-wave method  using our own
code FLAIR\cite{Weinert1982,Weinert2009} and also by comparing to a recent
calculation with the code Wien2K\cite{pcwPB} (the later was also used by 
Karki \emph{et al.}\cite{Karki2010}).  All these calculations yielded the same
result for the band structure of Mo$_{3}$Al$_{2}$C$_1$ in the cubic $\beta$-Mn
type crystal structure, i.e., the one shown in panel (a) of Fig. \ref{fig6}.  

We have calculated the band structures both in a scalar relativistic
approximation, omitting spin-orbit coupling, and fully relativistically,
including spin-orbit coupling in a self-consistent manner.  As expected, some
degeneracies at the high-symmetry points are different whether spin-orbit
coupling is included or not, e.g., a splitting of $30~$meV at $\Gamma$ and of $53~$meV at R in
Mo$_{3}$Al$_{2}$C$_{1}$ and of $20~$meV at both $\Gamma$ and R in
Mo$_{3}$Al$_{2}$C$_{0.75}$ just below the Fermi energy $E_\text{F}$ is evident.  However,
the most striking point is the loss of the double degeneracy of the bands due
to spin-orbit coupling in noncentrosymmetric
compounds.\cite{Callaway1964,Bauer2009} From Fig. \ref{fig6} we observe
this vertical spin-orbit splitting of the bands, e.g., $65~$meV on
the path $\Lambda$ around $-0.35~$eV in Mo$_{3}$Al$_{2}$C$_{1}$ and $90~$meV on
the path $\Lambda^*$ around $-0.3~$eV in Mo$_{3}$Al$_{2}$C$_{0.75}$.
As a consequence the Fermi surfaces of noncentrosymmetric Mo$_{3}$Al$_{2}$C$_{1-x}$
do also split due to spin-orbit coupling which can be seen in Fig. \ref{fig6} as the horizontal 
band splitting at $E_\text{F}$.  

Comparing the band structure and electronic DOS of 
Mo$_{3}$Al$_{2}$C$_{1}$ with that of Mo$_{3}$Al$_{2}$C$_{0.75}$ in Fig.
\ref{fig6} one immediately notices that removing a carbon atom from the carbon
sublattice has a substantial effect and it is not sufficient to simply 
shift the Fermi level in order to account for the carbon vacancy. 
  
From these results we conclude, that the structure of the Fermi surfaces and
hence any nesting, paramount to the understanding of superconductivity,
strongly depend on $x$.      

\section{Conclusions}

In the present work we have shown, that vacancies are necessary to dynamically stabilize
the cubic $\beta$-Mn type (P4$_1$32) crystal structure of Mo$_{3}$Al$_{2}$C,
whereby vacancies on the carbon sublattice are energetically the most
favorable ones.  According to our thermodynamical model the most probable carbon
vacancy concentration $x$ in Mo$_{3}$Al$_{2}$C$_{1-x}$ is about $0.13-0.14$
considering actual experimental preparation temperatures.

We have demonstrated that there exists a critical value of $x_\text{crit}\sim0.09$
below which Mo$_{3}$Al$_{2}$C$_{1-x}$ becomes dynamically unstable, and especially
the frequency at which the Debye-like behavior of the phonons ends strongly
depends on $x$. 

Likewise, the band structure and electronic DOS are influenced by the carbon
vacancy concentration.  

The still unresolved question if Mo$_{3}$Al$_{2}$C is a {\em conventional} or
{\em unconventional} superconductor can only be answered when the carbon
vacancies are properly considered, as the structure and nesting of the Fermi
surfaces depend on the carbon vacancy concentration.  If this carbon vacancy
concentration could be controlled, it might be possible to tune the
superconducting properties of Mo$_{3}$Al$_{2}$C$_{1-x}$.  This might be rather
difficult, because it may depend on the sample preparation and the cooling
process.  Further, one can safely assume, that at the preparation temperature
of $\approx 1500~$K the sample is in its thermodynamic equilibrium. This is not the case
when its superconducting properties are measured where it is in a quenched
meta-stable state. 

\acknowledgments
This work was supported by the Austrian Science Fund FWF under Grant No. P22295.
Computational calculations were done on the Vienna Scientific Cluster (VSC).  
\bibliography{citations}
\end{document}